\documentclass[useAMS,usenatbib,epsfig, a4paper]{mn2e}
\usepackage{times}
\usepackage{amsmath, amssymb}
\usepackage[dvips]{graphicx}

\newcommand{\be}{\begin{equation}}
\newcommand{\ee}{\end{equation}} 
\newcommand{\eei}{\end{equation}\indent\indent}
\newcommand{\bc}{\begin{center}}
\newcommand{\ec}{\end{center}}
\newcommand{\ber}{\begin{eqnarray}}
\newcommand{\eer}{\end{eqnarray}}
\newcommand{\ba}{\begin{array}}
\newcommand{\ea}{\end{array}}

\newcommand{\ls}{{\cal L}}

\newcommand{\lsim}{\,\raise 0.4ex\hbox{$<$}\kern -0.8em\lower 0.62ex\hbox{$\sim$}\,}
\newcommand{\gsim}{\,\raise 0.4ex\hbox{$>$}\kern -0.7em\lower 0.62ex\hbox{$\sim$}\,}

\def\case#1/#2{\textstyle\frac{#1}{#2} }
\newcommand{\bra}[1]{\left(#1\right)}
\newcommand{\bras}[1]{\left[#1\right]}

\title{Using the Topology of Large Scale Structure to constrain Dark Energy}
\author[Zunckel, Gott \& Lunnan ]{Caroline~Zunckel$^{1,2}$, 
J.  Richard Gott III$^{1}$ and Ragnhild Lunnan$^{3}$\\
$^{1}$ Astrophysics Department, Princeton University, Peyton Hall, 4 Ivy Lane, NJ, 08544, USA\\
$^{2}$ Astrophysics and Cosmology Research Unit, University of Kwazulu-Natal, Westville, Durban 4000, South Africa\\
$^3$ Harvard-Smithsonian Center for Astrophysics, 60 Garden St., Cambridge MA 02138, USA}
\date{\today}
\begin{document}
\maketitle

\begin{abstract}
The use of standard rulers, such as the scale of the Baryonic Acoustic oscillations (BAO), 
has become one of the more powerful techniques employed in cosmology to probe the 
entity driving the accelerating expansion of the Universe.  
In this paper, the topology of large scale structure 
(LSS) is used as one such standard ruler to study this mysterious `dark energy'.  
By following the redshift evolution of the clustering of luminous red galaxies (LRGs) as measured by their 3D topology (counting structures in the cosmic
web), we can chart the expansion rate and extract information about the equation of state of dark energy. 
Using the technique first introduced in (\cite{PK}), we evaluate the constraints that can be 
achieved using 3D topology measurements from next-generation LSS surveys such as the Baryonic Oscillation Spectroscopic Survey (BOSS).  
In conjunction with the information that will be available from the Planck satellite, 
we find a single topology measurement on 3 different scales is capable of constraining a single dark energy parameter 
to within $5\%$ and $10\%$ when dynamics are permitted. This offers an alternative use of the data available from redshift surveys and
serves as a cross-check for BAO studies.   
\end{abstract}

\section{Introduction}
In the late 90's, the expansion of the Universe, first detected by Hubble,
was confirmed from the precise measurements of SNIa and astonishingly found to be accelerating 
(\cite{Riess1998,Perlmutter:1998np,Lange:2000iq,Hoekstra:2002xs,Riess:2004nr,Cole:2005sx,Astier:2005qq,WMAP3,Riess:2006fw}).  
This provided convincing evidence for the presence of an unidentified entity in 
the Universe which (in the context of General Relativity) must act against the 
gravitational attraction of ordinary matter.  Indirect yet also compelling 
evidence came from the missing energy density inferred from the discrepancy between the 
measurements of matter density (from direct measurements and observation of the Integrated Sachs-Wolfe effect) 
and the indications of spatial flatness from the CMB anisotropy spectrum, 
as well as the level of the initial inhomogeneity measured in the CMB compared with large scale structure today. 
Although the presence of this `dark energy' is now well-established, its nature still evades us and characterizing it has become 
one of the most important topics in cosmology today.   This is evidenced by the large number of experiments which have been 
proposed and designed with this question in mind.

One such effort is the Baryon Oscillations Spectroscopic Survey (BOSS) which plans to map the spatial distribution of 
luminous red galaxies (LRG) and quasars over $10,000$ sq. deg. of the sky. The survey hopes to detect the excess of 
galaxy clustering at $100$ Mpc/h separations left over from the acoustic oscillations in the 
baryon distribution at the time of last scattering.  The change in the characteristic scale of this 
phenonemon from the time of the CMB to today is encapsulated by the diameter angular distance $d_A(z)=(1+z)^{-1} r(z)$, 
which is related to the expansion rate of space via the comoving distance
\be
r(z) = c \int^{z}_0 \frac{dz'}{H(z')}.
\label{rz}
\ee 
where 
\be
H(z) = H_0 \sqrt{\Omega_m\bra{1+z}^3+\Omega_{X} \exp \bra{\int_0^z \frac{1+w(z')}{1+z'} dz'}}
\label{w}
\ee where $H_0$ is the Hubble parameter today (note that flatness is assumed), $c$ is the speed of light, 
$\Omega_m$ is the current matter density and $\Omega_X$ is the current density of dark energy. We focussing on measuring 
the equation of state $w(z)$ of the dark energy component, which describes the
ratio of its pressure to its energy density as a function of redshift.
Because the scale of the oscillations at the time of last scattering is measured 
precisely from the CMB peak morphology, the BAO scale becomes a standard ruler. 
BOSS is forecasted to measure $d_A(z)$ to $1\%$ at various redshifts (\cite{BOSS}), 
placing constraints on the equation of state of dark energy $w$ in Eqn.~ \ref{w}.  
In this paper, we will use the topology of large scale structure as another such standard 
ruler with which to get a handle on $w(z)$.  Being a tracer of the primordial density 
perturbations, LSS provides a record of the initial conditions and so its topology has 
been used extensively to test our current assumptions of the earliest epoch. For example, the measured topology of
the Sloan Sky Digitial Survey (SDSS) traced by LRGs was shown to be consistent with expectations from a Universe with 
Gaussian randon phase initial conditions (\cite{Gott_SDSS}).  In this paper, we recognize that topology is another 
measure of clustering or the number of 
structures per unit volume and thus by mapping how galaxy clustering per unit volume evolves with 
redshift, the expansion history of the Universe can be studied. 
The potential of this dataset in dark energy studies was first recognized in (\cite{PK}), in which the genus statistic 
used to characterize LSS is adapted to map $r(z)$ at various epochs, thereby probing $w(z)$.  
Using the method presented in (\cite{PK}), we build on this analysis with the aim of determining how effective 
the measurement of topology from future LSS surveys such as BOSS will be 
at elucidating the nature of dark energy.  We find that the topology is indeed effective in placing constraints
of the average equation of state (at the $5\%$ level). The constraints do however weaking when a smoothly evolving  equation of state is 
considered, with very little information on a second dark energy parameter delivered.  
In Section \ref{genus_stats}, the genus statistics are reviewed, followed by a discussion of 
their application in Cosmology in Section \ref{cosmo}. Section \ref{analysis} gives the details of the analysis. In Section \ref{results}, 
the method is applied to the BOSS data as well as
information about the matter density and Hubble parameter from the Planck 
satellite.

\section{The Genus and related Statistics}
\label{genus_stats}
 
Because the perturbations in the primordial density field on scales larger than the 
correlation scale underwent linear growth, the pattern of matter overdensities 
today should reflect the distribution of these seeds from which they formed - - 
high density regions such as galaxies and clusters of galaxies, are in fact amplifications 
of the primordial density perturbations. This means that the distribution of the structure 
today on large scales gives us a window to the conditions of the Universe in a much 
younger state. 
As pointed out in \cite{GMD}, this initial state of the perturbations is not recorded 
only in the pattern of the overdense regions; the cosmic underdensities also form 
part of entire structure and it is the cosmic sponge on scales larger than the 
RMS displacement of the matter that remains preserved (\cite{PK}).  The topology 
of LSS can therefore be used to directly test predictions of 
our current theories describing the initial conditions.  To measure the topology, the 
number density distribution is smoothed with a Gaussian smoothing ball of radius 
$R_g$. We then find the iso-density contours of the smoothed distribution which 
divides the space into two, where the fraction volume on the high density side given by
\be
f(\nu) = \frac{1}{\sqrt{2\pi}} \int^{\infty}_{\nu} e^{-x^2/2} dx .
\ee
Here $\nu$ is merely a label for the contours. For example, $\nu=0$ denotes the 
iso-contour of the median density which encloses $f=50\%$ of the volume, while $\nu=1.5$ labels the $f=7\%$ contour.
One can then define the genus as a function of $\nu$, given by the difference between the number of donut-like holes and isolated regions;
\be
G(\nu) = \text{No. of holes} - \text{No. of isolated regions} .
\ee 
For a random Gaussian phase density field, the genus per unit volume, $g\bra{\nu} = G(\nu)/V$, is predicted from theory to be (c.f. \cite{HGW,GWM})
\be
g(\nu) = A \bra{1 - \nu^2} e^{-\nu^2/2}
\label{eq:g}
\ee
where the amplitude is given by
\be
A = \frac{\bra{\langle k^2\rangle/3}^{3/2}}{2\pi^2} .
\ee  Here $\langle k^2\rangle$ is the average value of $k^2$ in the 
smoothed power spectrum. $A$ is sensitive to the slope of the power 
spectrum near the selected smoothing scale and is independent of the 
amplitude of $P(k)$.  Recent studies of the topology of LSS traced by LRGs in the SDSS survey \cite{Gott_SDSS}
at smoothing lengths of $21$~h$^{-1}$Mpc and $34$~h$^{-1}$Mpc show the amplitude of the genus curve, $A$, is modeled very well
by N-body simulations and closely follows that of the initial conditions. Note that the topology of a non-Gaussian density 
field at any scale within the linear regime is also preserved in comoving space. 

The currently favoured model for the beginnings of structure formation, called Inflation, 
assumes that the primordial fluctuations to be Gaussian random 
phase, which has been shown to lead to a sponge-like topology (\cite{GWM, GMD}). 
The medium density contours measured for various LSS samples over the past few 
years have been found to be consistent with expectations of this type of topology 
(see \cite{Gott_SDSS,PKG,Park,Gott,Gott_1989,HGW}). 
In addition, the theoretical prediction for Gaussian random 
phase initial conditions in Eqn. \ref{eq:g} was shown to provide a 
suitable fit to the observed genus curves for these data sets.  
In this paper, we focus of whether the same data can be used to 
study the late-time behavior of the Universe.

\section{Using Topology to constrain Cosmology}
\label{cosmo}

The comoving distance $r(z)$ in Eqn.~ \ref{rz} tells us how scales change 
as a function of redshift due to the expansion of space. This means that 
comparison of characteristic scales in cosmology, such as features in the 
power spectrum or the correlation function, from one epoch to the another, 
provides a handle on $r(z)$.  
Because the shape of the primordial power spectrum $P(k)$ on large scales (where we are still in the linear regime) 
should be conserved in redshift space (the transfer function does not depend on $z$), we can determine $P(k)$ 
for a given set of cosmological parameters from two measurements of $P(k,z)$ and 
expect to find it unchanged. If the power spectrum does change, the $r(z)$ 
relation in \ref{rz} and hence the chosen cosmology must be wrong.  

As described in Section \ref{genus_stats}, the genus statistic is related 
to the entire shape of the power spectrum and essentially measures its slope 
near the smoothing scale. The statistic $g(\nu)$ is a measure of the number 
of structures after smoothing on a certain scale, per unit volume, out to a 
given distance. The measure therefore relies on a choice of $r(z)$ relation 
(and hence cosmology) to compute the enclosed volume.   Assuming the incorrect 
cosmology and hence expansion rate will lead to an incorrect estimate of the 
volume in which one is counting structures. In so doing, a different amplitude 
of the genus curve from the true value will be measured.  Furthermore, the 
chosen smoothing scale which determines the size of the structures will be 
incorrect.  For example, in the case where one overestimates the expansion rate, 
what is assumed to be a unit volume will enclose less structure than at $z=0$.  
As a result, the 
measured amplitude of structure on the smoothing scale will be lower.  At the 
same time, because the smoothing scale per unit volume is kept fixed, smaller 
structures are effectively smoothed over so these two effects partially cancel. 
Fortunately, there is a net effect because the density perturbation field is 
not scale free and the power spectrum has a shape with a varying slope over 
the region of interest. Fig.~ \ref{g} shows the genus curves for two different cosmological models. 
In this paper, we will exploit the dependence of the genus curve on $r(z)$ in 
order to constrain the equation of state of dark energy, as explained in the next section.

We consider a flat Universe filled with cold dark matter, baryons and dark energy. 
The cosmological parameters of interest are the matter density, $\Omega_m$ (baryonic and dark), 
the Hubble parameter $H_0$ and the equation of state of the dark energy component, 
which has been shown to provide a suitable phenomenological description of dark energy (\cite{review}).  
To start, we assume the simplest parameterization, namely a constant equation of state $w$.  
The comoving distance along the line of sight, $r_{||}$ (in redshift
space), and in the traverse direction, $r_{\perp}$, of a feature
sitting at a redshift $z$, are related to the redshift range
$\Delta z$ covered and the angle subtended $\Delta \Theta$
respectively, by
  \be
 r_{||}=c\frac{ \Delta
z }{H(z)}, \quad r_{\perp}=c(1+z)d_{A}(z) \Delta \Theta.
\label{eq:BAO}
 \ee

Suppose a length along the line of sight spans a redshift range of $\Delta z$. The comoving distance of the length is given by
\be
\lambda_x = r_{\mid\mid}(z) = \frac{c \Delta z}{H\bra{z, \Omega_{m}, w, H_0}}. 
\ee If we then perturb the cosmology, the same redshift range $\Delta z$ will correspond to a different comoving distance given by
\be
\lambda_x' = \frac{c \Delta z}{H\bra{z, \Omega_{m}', w', H_0'}}. 
\ee
Thus the different comoving distances are related by
\be
\lambda_x' =\lambda_x \frac{H\bra{z, \Omega_{m}, w, H_0}}{H\bra{z, \Omega_{m}', w', H_0'}}.
\ee

The comoving distance subtended by an angle $\Delta \Theta$ is given by
\be
\lambda_z = r_{\perp}(z) = (1+z) d_A\bra{z, \Omega_{m}, w, H_0} \Delta \Theta
\ee 
If we again perturb the cosmology, the same angle $\Delta \Theta$ will subtend a new comoving distance of 
\be
\lambda_z' = (1+z) d_A\bra{z, \Omega_{m}', w', H_0'} \Delta \Theta
\ee 
So we can write
\ber
\lambda_z' &=& \lambda_z \frac{(1+z)}{(1+z)} \frac{d_A\bra{z, \Omega_{m}', w', H_0'}}{d_A\bra{z, \Omega_{m}, w, H_0}}\nonumber\\
&=& \lambda_z \frac{d_A\bra{z, \Omega_{m}', w', H_0'}}{d_A\bra{z, \Omega_{m}, w, H_0}}. 
\eer

Similarily, the comoving distances subtended by the angle $\Delta \Theta$ in the $y$ direction in two different cosmologies are related by
\be
\lambda_y' =  \lambda_y \frac{d_A\bra{z, \Omega_{m}', w', H_0'}}{d_A\bra{z, \Omega_{m}, w, H_0}}
\ee
We can then write the volume of the corresponding ellipsoid
\be
V = \frac{4\pi}{3}\lambda_x \lambda_y \lambda_z
\ee so we can relate the comoving volumes in the two different cosmologies via

\ber
V' &=& \frac{4\pi}{3}\lambda_x' \lambda_y' \lambda_z'\nonumber\\
&&\bras{\lambda_z \frac{d_A\bra{z, \Omega_{m}', w', H_0'}}{d_A\bra{z, \Omega_{m}, w, H_0}}}\nonumber\\
&=&\frac{4\pi}{3}(\lambda_x\lambda_y \lambda_z) \frac{H\bra{z, \Omega_{m}, w, H_0}}{H\bra{z, \Omega_{m}', w', H_0'}}\bras{\frac{d_A\bra{z, \Omega_{m}', w', H_0'}}{d_A\bra{z, \Omega_{m}, w, H_0}}}^2\nonumber\\
&=& V \frac{H\bra{z, \Omega_{m}, w, H_0}}{d_A\bra{z, \Omega_{m}, w, H_0}^2} \frac{d_A\bra{z, \Omega_{m}', w', H_0'}^2}{H\bra{z, \Omega_{m}', w', H_0'}} 
\eer
so
\be
 \frac{V'}{\bras{\frac{d_A\bra{z, \Omega_{m}', w', H_0'}^2}{H\bra{z, \Omega_{m}', w', H_0'}}}}
 =  \frac{V}{\bras{\frac{d_A\bra{z, \Omega_{m}, w, H_0'}^2}{H\bra{z, \Omega_{m}, w, H_0}}}}.
\ee
We denote the factor relating the volumes in the different cosmologies as 
\be
V_{eff} = \frac{d_A\bra{z, \Omega_{m}, w, H_0}^2}{H\bra{z, \Omega_{m}, w, H_0} }.
\ee 
Re-arranging the above gives an equation which gives the volume in the new cosmology as a function of the original volume;
\be
 V'
 = V \frac{V'_{eff}}{V_{eff}}.
\ee

Suppose we have a galaxy redshift catalogue with which we 
wish to constrain the underlying cosmology. After smoothing the 
distribution with a Gaussian smoothing sphere of dimensions 
$\lambda = \lambda_x = \lambda_y = \lambda_z$ and volume $V = \lambda^3$,  
we measure an amplitude of $G = g(\lambda, z) V = g(\lambda, z) \lambda^3$. 
Now suppose that the assumed cosmology for the smoothing process is in fact incorrect. 
This means that the $r(z)$ relation is wrong. In the true cosmology, 
the smoothing sphere is in reality an ellipsoid with volume $V' = \lambda_x' \lambda_y' \lambda_z'$. 
The same amplitude $G = g' (\lambda', z) V'$ is measured by 
smoothing a different cosmology with an ellipsoid or a sphere of radius $\lambda'$, which is calculated using
\ber
\frac{\lambda'}{\lambda} &=& \bra{\frac{V'}{V}}^{1/3} \nonumber\\
&=&\bra{\frac{V_{eff}'}{V_{eff}}}^{1/3}. 
\eer

So we can relate the smoothing scales within the two different cosmologies using
\be
\lambda'=\lambda \bra{\frac{V_{eff}'}{V_{eff}}}^{1/3} .
\ee
 Equating $g V = g' V'$ gives us
\ber
g'\bra{\lambda', z} &=& g\bra{\lambda, z} \frac{V}{V'} \nonumber\\
&=& g\bra{\lambda, z}   \frac{V_{eff}}{V'_{eff}}.
\label{eq_g}
\eer

Eqn. \ref{eq_g} is very useful because it allows us to calculate 
the expected genus curve for any cosmological model provided we know the theoretical 
$g(\lambda, z)$ at a given scale $\lambda$ for one set of parameter values.  This is 
readily computed using the smoothed matter power spectrum generated using a Boltzmann 
code such as CAMB (\cite{CAMB}).  The validity of Eqn.~ \ref{eq_g}  depends on how well 
a galaxy distribution smoothed with a sphere of geometric mean $R_g$ represents that smoothed 
using an ellipsoid of median radius $R_g$. This is tested and discussed in the next section.

\section{Analysis}
\label{analysis}
The aim of the analysis is to perform dark energy forecasts for a set of measurements of the genus at various smoothing scales, 
based on the 3D distribution of LRGs as it will be measured by BOSS.  Using a $2048^3$ particle cold dark matter N-body simulation, 
\cite{PK06}'s technique which identifies LRG galaxies by selecting the most massive bound halos, 
correctly reproduces the 3D topology of the LRG galaxies in the SDSS \cite{Gott_SDSS}. 

We apply the same technique to the Horizon Run Simulation, which is the largest N-body simulation to date with a volume of $6.543$ (Gpc/h)$^3$ 
(\cite{KPGD}). Provided we remain in the linear regime, the genus curve for the initial conditions remains essentially
unchanged and should provide a suitable representation of the genus that would be measured 
using LRGs as a probe of LSS today. Hence we use the initial density field from the Horizon Run Simulation to construct a 
hypothetical data set and subdivide the simulation into 8 cubes. Placing an observer at the centre of each cube, a sphere of 
comoving radius $1570$ Mpc/h (corresponding to a redshift of $z=0.6$) is carved around the observer and 
quartered to produce 4 BOSS mock catalogues of $\pi$ steradians and in total, 32 mock galaxy 
catalogues with which to test the statistics.  The genus per unit smoothing volume is then measured for 
each catalogue at 3 different smoothing scales; $15, 21, 34 ~\text{Mpc h}^{-1}$.  These scales were selected 
because they are sufficiently disparate that different types of structures are being smoothed and can, 
for statistical purposes, be regarded as independent.  This gives three independent data points 
$g_{data}(z,\lambda_i)$ where $i=1-N_z$ per redshift bin.  The noise is taken to be the standard 
deviation of the measured $g$ values of the 32 simulated maps. 

\begin{figure}
\begin{center}
\begin{tabular}{c}
    \includegraphics[trim = 0mm 0mm 0mm 0mm, scale=0.7, angle=0]{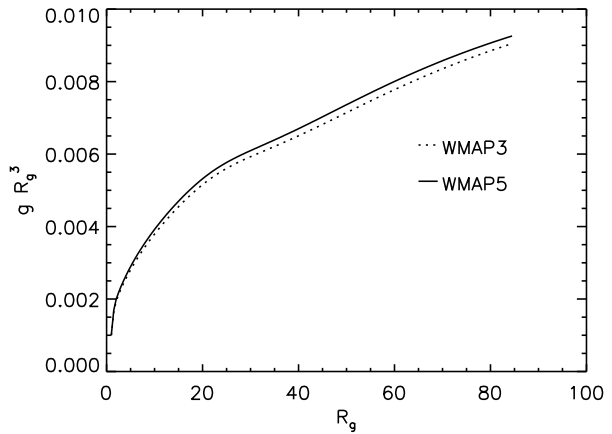}
\end{tabular}
  \end{center}
\caption{Plot of the amplitude of the genus curve $g R_g^3$ as a function of the smoothing length $R_g$ for the best-fit parameters 
values for WMAP 3-year ($\Omega_m=0.24, H_0=72$) and WMAP 5-year data ($\Omega_m=0.26, H_0=74$) 
assuming a flat $\Lambda$CDM cosmology.}
\label{g}
\end{figure}

First, we compare the genus per unit smoothing volume measured in our mock surveys to the 
theoretical values from Fig.~\ref{g}. As it should, the shape of the genus curves follows 
that of Gaussian random fluctuations (Eqn.~\ref{eq:g}), but we systematically find slightly 
lower amplitudes than predicted from theory. In particular, the suppression is systematically 
larger with larger smoothing length, and in redshift bins with more surface pixels. We believe 
this is due to two related effects: pixels on the edges are not counted by the program measuring 
the genus, and structures near the edges are supressed by the smoothing and ``cut off'', 
both lowering the value of the genus measured. To account for this, we introduce correction 
factors for each redshift bin, so that for the original density field, the mean values for the 
given bin correspond to the theoretical value. 

Next, to test the validity of Eqn.~\ref{eq_g}, we remap the density field in each survey, according to 
two different values of constant $w$: $-0.9$ and $-0.8$. For each cosmology, we construct a new density field 
from the original, based on the distances the observers would infer according to those cosmologies.
 We then again compute the genus per unit volume for the same three smoothing lengths and redshift bins, 
in our 32 mock surveys. Each bin is corrected for edge effects by the factors calculated above, and then
 compared to the expected number based on the geometric mean volume (Eqn.~\ref{eq_g}). We find that for 
both cosmologies, in all of the redshifts bins, for all smoothing lengths, the differences between 
the observed values in the simulated cosmologies (with $w=-0.8$ and $w=-0.9$) and 
theoretical values using Eqn. ~\ref{eq_g} are well within one standard deviation of the mean. We conclude that as 
long as edge effects are taken into account, the theoretical value based on effective smoothing length (Eqn.~\ref{eq_g}) 
agrees excellently with the observations. Therefore, we go ahead and use Eqn.~\ref{eq_g} to 
compute genus curve amplitudes in different cosmologies for the likelihood analysis.

We now wish to evaluate the likelihood of the parameters $\bar{p} = \bra{H_0, \Omega_m, w}$ in light of the data. This involves computing 
the expected values of the genus per unit volume at the point $\bar{p}_i$ in parameter space using Eqn.~ \ref{eq_g} for a given smoothing length and 
comparing them with the data using
\be
\chi^2 = \sum_{i=j,N_z}\sum_{i=1,3} \frac{\bra{g_{data}(z_j, \lambda_i) - g_{trial}(z_j, \lambda_i)}^2}{ \sigma_{ij}^2}
\ee
where
\ber
g_{trial}\bra{\lambda, z} &=& g_{ref}\bra{\lambda', z}   \frac{V^{ref}_{eff}}{V^{trial}_{eff}}.
\eer
and $\sigma_ij$ is the uncertainty associated with the data smoothed on the i$^{th}$ scale in the j$^{th}$ redshift bin. 
This is computed from variance measured from the 32 mock surveys.  The likelihood of the trial point is then $\ls\bra{\bar{p}_i| g_{data}}$.

In the above, $g_{ref}\bra{\lambda, z}$ are the expected genus values for our reference cosmology, 
calculated from the linear matter power spectrum generated using CAMB and smoothed on the scale
\be
\lambda' = \lambda \bra{\frac{V^{ref}_{eff}}{V^{trial}_{eff}}}^{1/3} .
\ee
The value of $g_{ref}\bra{\lambda', z}$ at this new smoothing scale $\lambda'$ is found using 
interpolation. We use the WMAP 3-year best-fit cosmological parameters to generate $g_{ref}\bra{\lambda, z}$ 
shown in Fig.~ \ref{g}.   This procedure is repeated at each point in parameter space, which is efficiently 
sampled using a Monte Carlo Markov chain (MCMC) algorithm until convergence. 
The $68\%$ and $95\%$ confidence intervals for the parameters $\bra{H_0, \Omega_{X}, w}$ are then computed from the chains.

\section{Results}
\label{results}

Assuming a completed BOSS survey, we constructed a hypothetical data set consisting of three genus measurements, each at one of the three
selected smoothing scales. We pick a dark energy model with $w\neq -1$ to test the ability of the data
to distinguish such a model from cosmological constant. 
We simulate a flat Universe with a dark energy component described by $w=-0.9$.

\begin{figure}
\begin{center}
\begin{tabular}{c}
    \includegraphics[trim = 0mm 0mm 0mm 0mm, scale=0.7, angle=0]{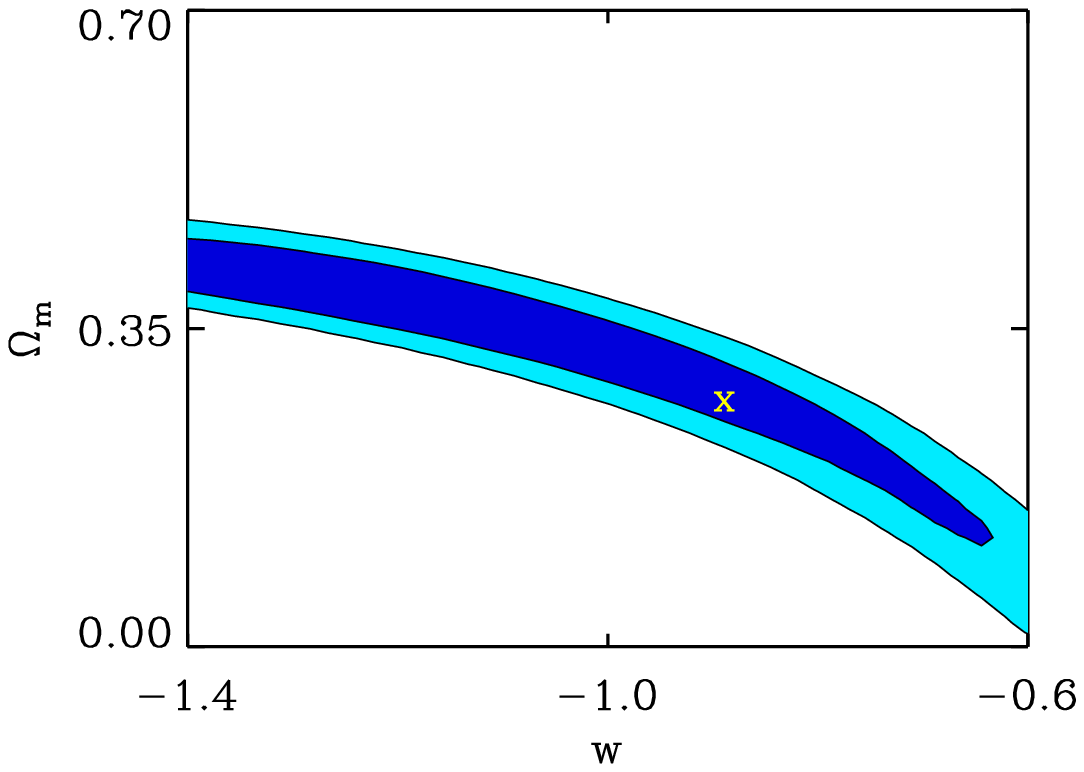} \\
(a)\\
\includegraphics[trim = 0mm 0mm 0mm 0mm, scale=0.7, angle=0]{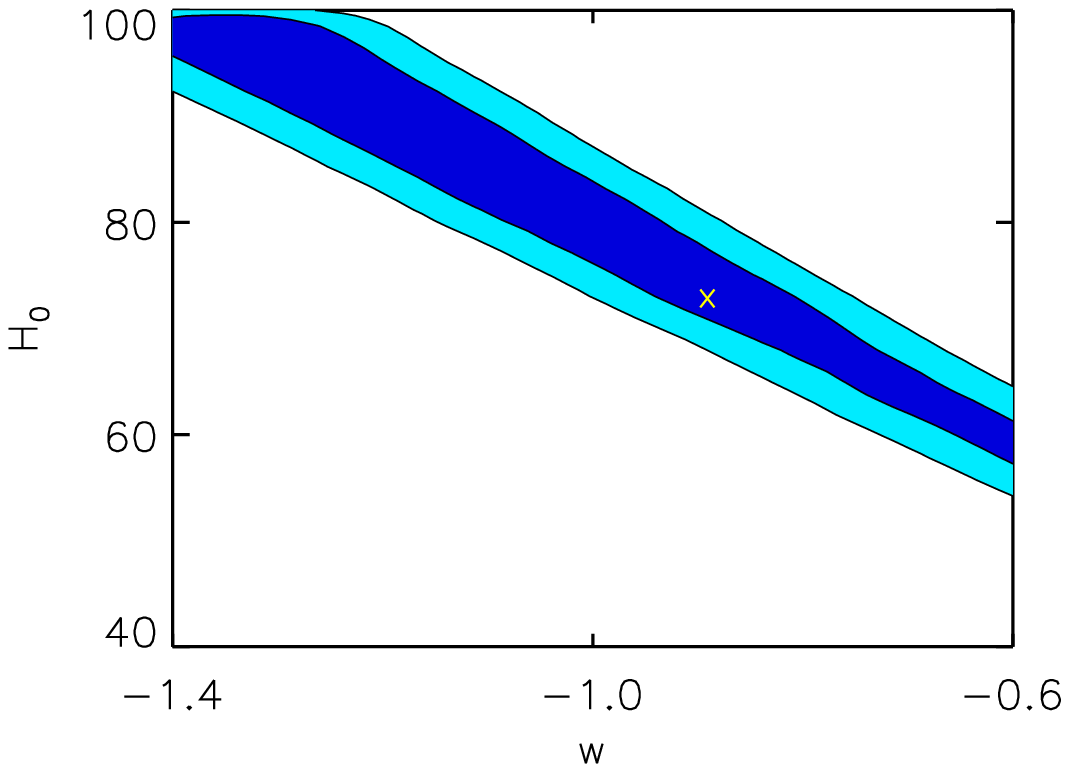}\\
(b)
\end{tabular}
  \end{center}
\caption{95$\%$ confidence regions (a) for a parameter space with $\bra{\Omega_{m}, w}$ 
where $H_0$ has been fixed and (b) for a parameter space $\bra{w, H_0}$ where $\Omega_m$ has been fixed. The true values of the parameters
used to produce the simulated topology data are marked by a yellow cross.  }
\label{fig1}
\end{figure}

Fig.~ \ref{fig1} shows the $68\%$ and $95\%$ confidence regions in parameter space 
when (a) $H_0$ is fixed while $\Omega_m$ and $w$ are 
allowed to vary and (b) $\Omega_m$ is fixed while $H_0$ and $w$ are allowed to vary. 
The elongated contours exhibit strong deneracy between the matter density and Hubble parameter and the equation of state. 
For a more negative choice of $w$ than the true value, the dark energy component is less 
important in the past while the matter density today is held fixed. 
In order to match the same genus measurement, a higher value of the Hubble constant 
today is needed in order to restore $H(z)$ and to yield the correct measurement of the 
sampled volume.  For a more positive choice of $w$ and fixed $H_0$, $H(z)$ is higher, 
indicating that space is expanding more rapidly. The implication is that the estimated 
volume out to a given $z$ will be smaller.  To counteract this, the matter density can be 
reduced  thereby increasing the contribution from the dark energy component to restore the 
expansion rate to a lower value.

In addition to the simulated data from BOSS, we choose to include information 
from the Cosmic Miscrowave Background (CMB). The presence of dark energy impacts the CMB primarily 
through the distance to the surface of last scatter, with the expansion rate dictating the 
scales on which the acoustic features appear in the CMB spectrum today.   The size of the 
largest full acoustic compression which manifests as the first peak in the CMB angular power 
spectrum, is determined solely by the distance that these waves could have traveled in the time before 
recombination, namely the sound horizon $r_{s}$ at last scattering;
\be
r_s \bra{z_{cmb}, \Omega_b, \Omega_r} = \int_0^{t_{cmb}} c_s dt 
\ee
where $c_s$ is the sound speed defined by
\be
c_s = c\bras{ 3\bra{1+\frac{3\Omega_b}{4\Omega_r}}^{-1/2}}
\ee
Since the baryon to photon ratio can be precisely measured 
from the acoustic peak morphology in the CMB,
we can predict $c_s$ and thus the scale of the first peak. 
The angular scale on which this feature appears today, $\theta_s$ depends on the expansion 
history and the distance to the surface of last scattering, $d_{sls}$;
\be
\theta_s = \frac{180^\circ}{\pi} \frac{r_s}{d_{sls}}
\ee
where
\be
d_{sls} = c\int^{z_{cmb}}_0 \frac{dz'}{H(z')}.
\ee
Assuming parameter values near the $\Lambda$ concordance model, we use the fitting formula
\be
r_s = 144.4  \text{Mpc}\bra{\frac{\Omega_b h^2}{0.024}}^{-0.252}  \bra{\frac{\Omega_m h^2}{0.14}}^{-0.083}.
\ee 
This assumes that energy density contribution from the dark energy 
component during this epoch is sufficiently small such that it does not drastically affect 
the size of the acoustic features at the surface of last scattering. The information regarding the expansion rate and hence dark energy available 
in the CMB data comes primarily from the distance to the last scattering surface $d_{sls}$. 

The current constraints from the WMAP 7-year data are $\theta_s = 0.597 \pm 0.0016$ with the matter and baryon densities estimated to be
$\Omega_m h^2 =  0.1098 \pm 0.0058$ and $\Omega_b h^2 =  0.0250 \pm 6.3 \times 10^{-4}$ respectively (\cite{WMAP7}). 
We also include priors based on forecasts for the Planck CMB experiment which should be available at the time that the
BOSS survey is completed.  The higher resolution and sensitivity of the Planck Satellite 
lead to tighter forecasted constraints of $\theta_s = 0.597 \pm 3.1\times 10^{-4}$,
$\Omega_m h^2 =  0.1098 \pm 1.4 \times 10^{-3}$ and $\Omega_b h^2 =  0.0250 \pm 1.6 \times 10^{-4}$ (\cite{colombo}).    
As a result of the degeneracy between $w$ and $H_0$, permitting non-$\Lambda$ models and allowing $w$ to 
vary significant degrades the constraints on $H_0$.   
We thus include the current measurement of the Hubble 
constant from $H_0 = 74 \pm 3.6$km s$^{-1}$Mpc$^{-1}$ (\cite{riess2009}).

Including information from the CMB is highly advantageous as the direction of the 
degeneracy between $\Omega_m$ and $w$ is perpendicular to that in the topology data.  
As $w$ becomes less negative, its contribution to the overall density at earlier epoch increases.   
As a result, the expansion rate is reduced, with the features on the last scattering surface 
subtending smaller angles, manifesting as a shift in the acoustic peaks to lower $\ell$. 
This effect on the CMB anisotropy spectrum can however be countered by an increase in the 
matter density which delays recombination and leads to a larger sound horizon $r_s$.  
Fig. \ref{figpriors_planck_priors} shows the allowed regions in parameter space for the simulated BOSS topology data in 
conjuction with the WMAP 7-year (left) and the Planck  data (right). 
The change in the direction of the $\bar{w, \Omega_m}$ ellipse points to degeneracy breaking. 
There is neglible improvement on the dark energy EOS constraint from $w=-0.92^{+0.5}_{-0.6}$ to $w=-0.92^{+0.05}_{-0.05}$ 
when the CMB data set is updated from WMAP-7 year to the Planck data, despite the clear reduction in the confidence interval evident in
Fig. \ref{figpriors_planck_priors}. This indicates the the topology data is 
primarily responsible for the constraints on $w$. 

A useful task would be to determining the constraining power of this dataset as a function of redshift. 
This can be studied by performing principal component analysis, which essentially identifies the 
directions in the data in which the variation is maximal. These directions are captured by 
z-dependent eigenvectors, with the eigenvalue of mode representing how well it is constrained 
(ie. the strength of the information it represents). This will reveal where in redshift space 
this particular dataset is most sensitive to $w$. 
 
\begin{figure}
\begin{center}
\begin{tabular}{c}
    \includegraphics[trim = 0mm 0mm 0mm 0mm, scale=0.7, angle=0]{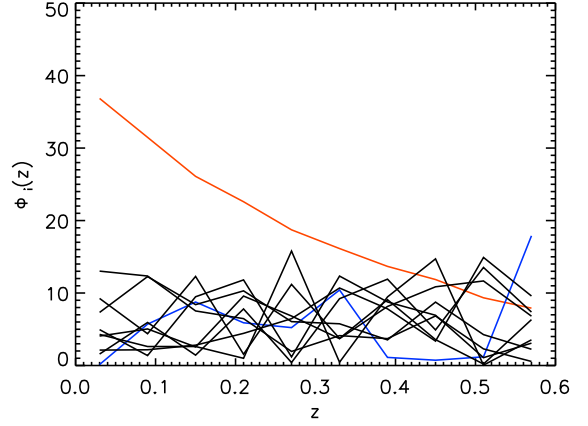} 
 \end{tabular}
  \end{center}
\caption{Figure showing a plot of $\phi_{i}(z)$ as a function of reshift. 
The first eigenvector $\phi_{1}(z)$ is shown in red, the second $\phi_{2}(z)$ in blue 
and with the remaining 8 shown in black. This illustrates the
redshift sensitivity of this dataset. }
\label{pca_eigenvecs}
\end{figure}

\begin{figure}
\begin{center}
\begin{tabular}{cc}
 \includegraphics[trim = 20mm 0mm 0mm 0mm, scale=0.4, angle=0]{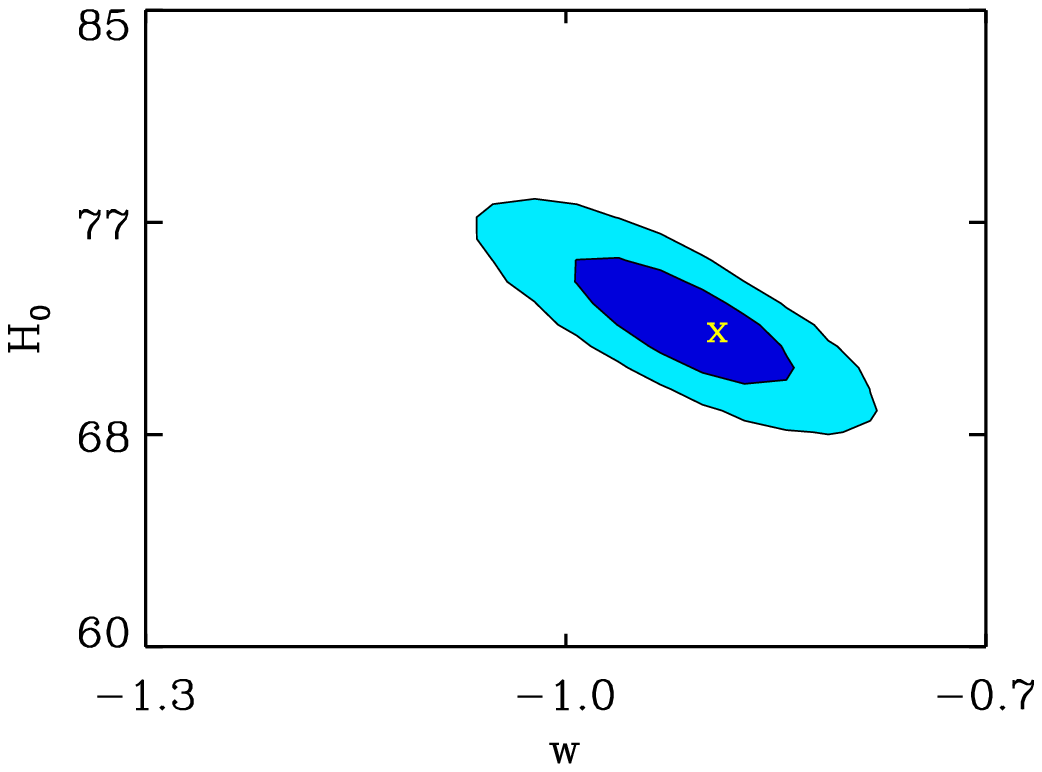} 
&   \includegraphics[trim = 20mm 0mm 0mm 0mm, scale=0.4, angle=0]{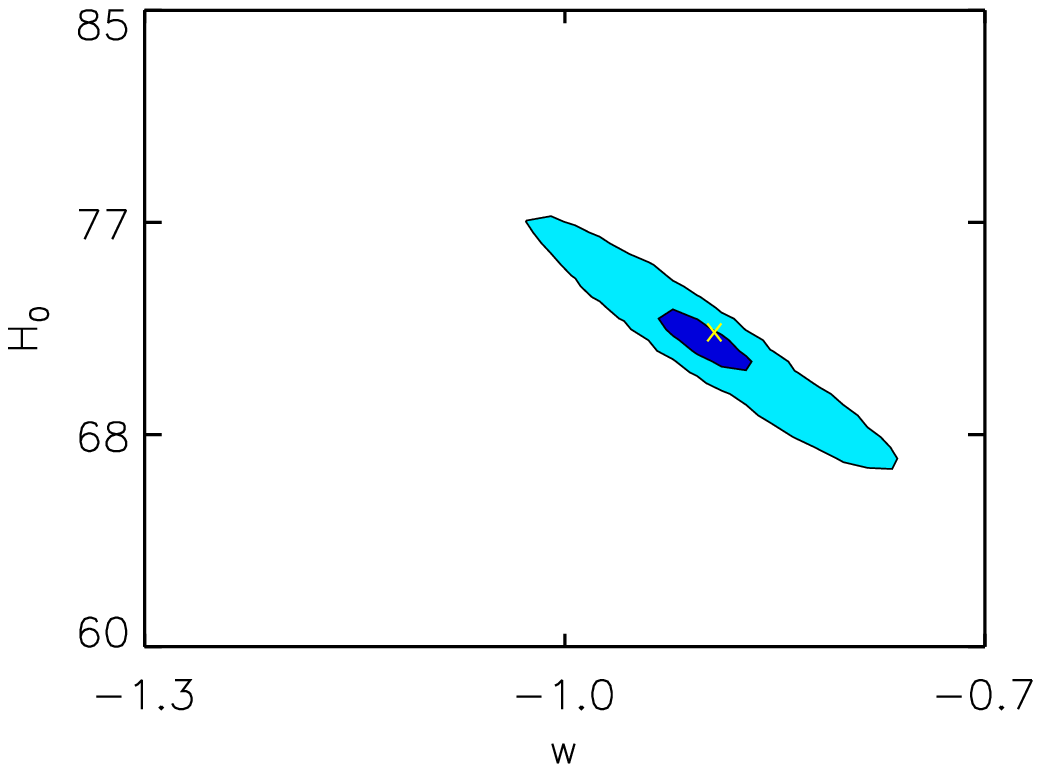}  \\
(a) & (b)\\
 \includegraphics[trim = 20mm 0mm 0mm 0mm, scale=0.4, angle=0]{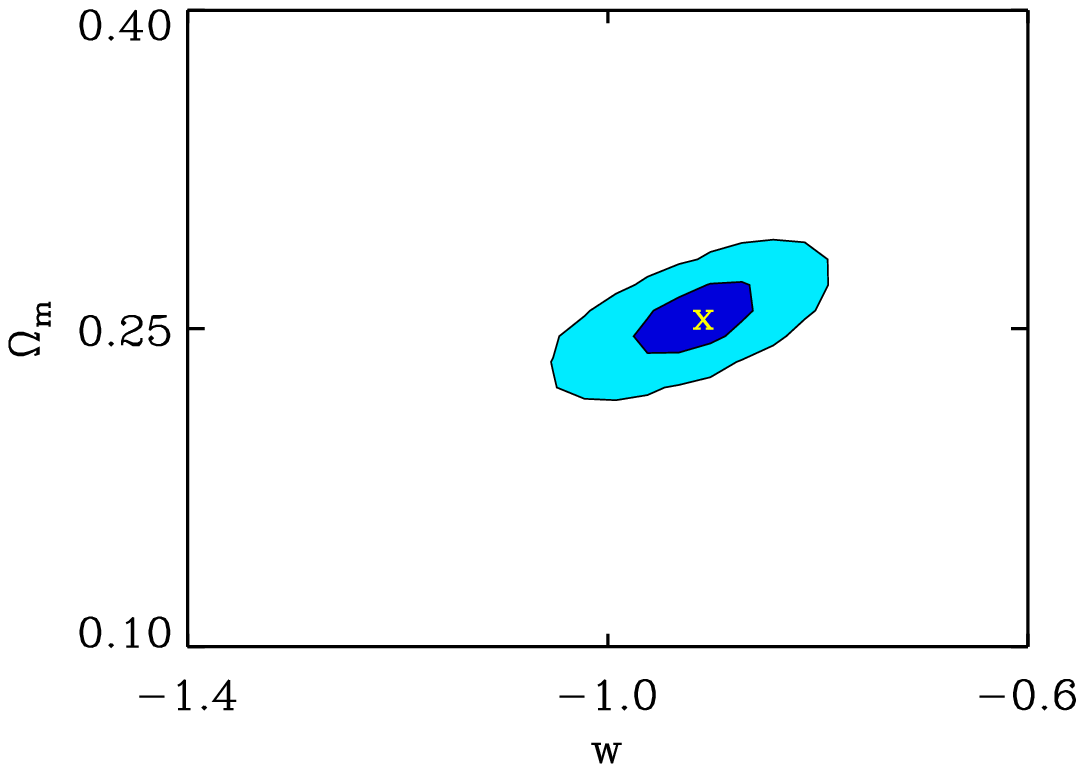} 
& \includegraphics[trim = 20mm 0mm 0mm 0mm, scale=0.4, angle=0]{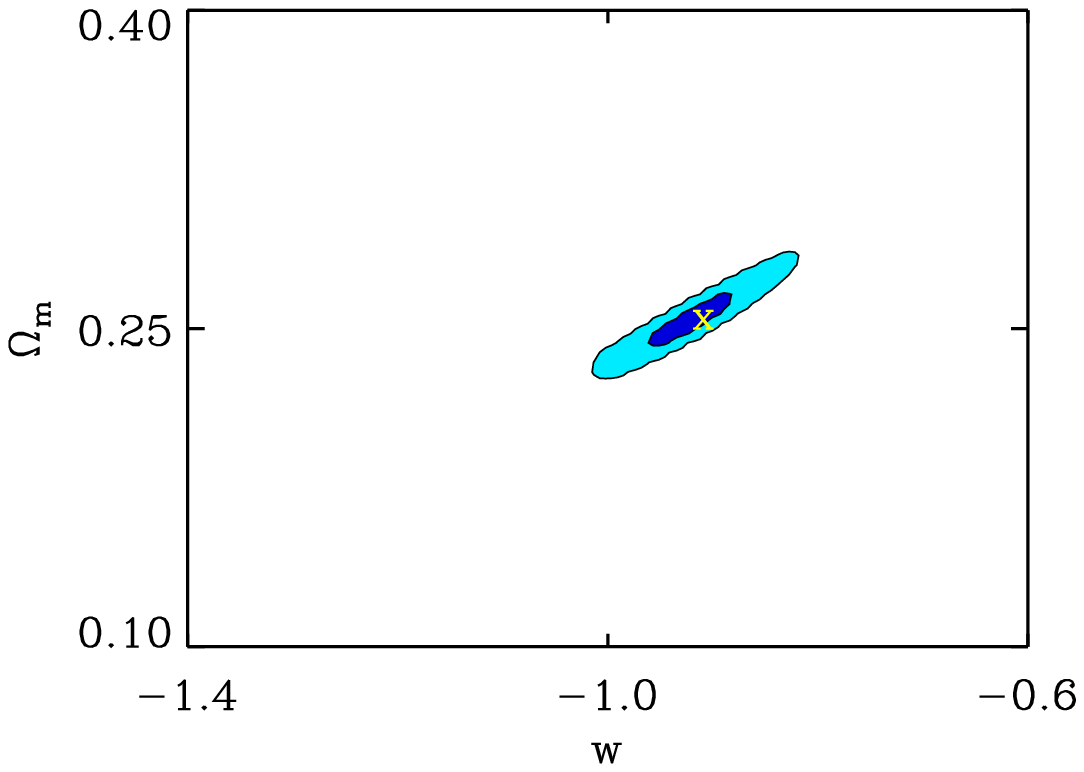}\\
(c) & (d)
 \end{tabular}
  \end{center}
\caption{95$\%$ confidence regions for $\bra{w, H_0}$ for the topology data including (a) WMAP 7 year data and (b) 
Planck data. Plots of $\bra{w, \Omega_m}$  for the same combinations of data sets are shown in (c) and 
(d). The true parameter values are shown by the yellow 
crosses. }
\label{figpriors_planck_priors}
\end{figure}

 We start by dividing the redshift region in 10 equally sized bins centered at redshifts $z_i$. 
 We parameterize the EOS as a piece-wise constant function, with a constant $w_i$ in each bin 
where $z_i-\Delta z/2<z_i<z_i+\Delta z/2$.  We then sample the posterior probability distribution 
an MCMC algorithm, marginalizing over the other cosmological parameters. Using the chain of samples, 
we construct the covariance matrix as follows;
\be
C_{ij} = \sum_{k=1}^N \frac{\bra{X_i - \bar{X_i}}\bra{X_j - \bar{X_j}}}{N-1}
\ee where $X_i$ is the value of the parameter at the i$^{th}$ point in the MCMC chain and 
$\bar{X}_i$ is the sample mean.  We find the eigenvectors by diagonalizing the covariance 
matrix and decomposing it as follows;
 \be
 C = W^T \Lambda W
 \ee where the diagonal matrix $\Lambda$ contains the eigenvalues corresponding to the 
eigenvectors ${\bf e}_i$ in the rows of matrix $W$.   We follow (\cite{Tang}) and track the redshift sensitivity of 
each mode by plotting the quantity $\phi_i(z)$;
\be
\phi_{i}(z) = N \mid\sqrt{\lambda_i}{\bf e}_i(z)\mid
\ee where the amplitude is proportional to the accuracy with which the eigenvector ${\bf e}_i(z)$ can be measured. 
$N$ is included to lessen the dependence on the number of bins.  Fig. \ref{pca_eigenvecs} plots $\phi_{i}(z)$ 
as a function of z.  The behaviour of the first and best-constrained mode (red) indicates 
that the constraining power of the survey is a decreasing function of $z$, with most of the 
information about $w$ delivered at low redshift.  All other eigenvectors are relatively 
noisy with no z-dependent shape distinct shape.  We infer that the redshift sensitivity 
of this data set is likely to behave as $\phi_1(z)$ (shown in red). 

\subsection{Redshift bins}
Up to this point we have smoothed the entire observed volume on 
three different scales, constructing a dataset comprising of three data points. 
In this Section we consider the effect of dividing the observed space into 
redshift bins of equal volume, thereby increasing the number of independent 
measurements three fold. The characteristics of the redshift shells in the case of one, 
two and three bins are summarized in table \ref{tab_zbins}. 

\begin{table}
\begin{tabular}{|cccc|}
	\hline
 Bins & Redshift range & Distance (Mpc/h) & Volume (Gpc/h)$^3$ \\
\hline
1 & $z < 0.6$ & $r < 1570$ & $4.036$  \\
\hline
2 & $z < 0.46$ & $r < 1246$ & $2.016$  \\
& $0.46 < z < 0.6$ & $1246 < r < 1570$ & $1.997$  \\
\hline
3 & $z < 0.396$ & $r < 1089$ & $1.342$  \\
& $0.396 < z < 0.513$ & $1089 < r < 1371$ & $1.326$  \\
& $0.513 < z < 0.6$ &  $1371 < r < 1570$ & $1.321$ \\
\hline
\end{tabular}
\label{tab_zbins}
\caption{Table summarizing the partitioning of redshift space.  }
\end{table}

The results of the analysis are summarized in the first section of table 
\ref{tab_zbins_results}. An increase in the number of redshift 
shells is shown to make little difference to the constraints when a constant equation of 
state is considered.  The increase in the number of data points is clearly 
countered by the reduction in signal-to-noise in each redshift shell over a single bin. In light of the finding that the sensitivity of the topology data to the dark energy model is a decreasing function of redshift  (shown in Fig. \ref{pca_eigenvecs}), it is likely that most of the information available about $w$ is concentrated in the first redshift bin, with the high-z bins adding little to the analysis.   

\subsection{Constraining dynamical dark energy}

\begin{figure}
\begin{center}
\begin{tabular}{c}
    \includegraphics[trim = 0mm 0mm 0mm 0mm, scale=0.7, angle=0]{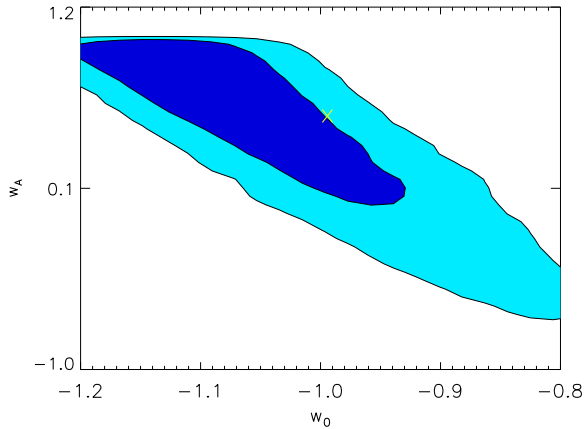} \\
 \end{tabular}
  \end{center}
\caption{ Plot of the 68$\%$ (dark blue)  and 95$\%$ (light blue) confidence regions for $\bra{w_0, w_a}$. 
The simulated data was divided into 3 equi-volume redshift bins summarized in table \ref{tab_zbins}. We include the Planck priors as well as the current constraint on the Hubble parameter.}
\label{figpriors_planck_priors_w0_wa}
\end{figure}

Up to this point, we have selected the simplest description of the 
equation of state, a constant $w$.  Although it has been shown to be a 
good approximation to a quintessense component obeying tracker solutions 
)\cite{efstathiou-1999}) while accommodating the vacuum energy ($w_0=-1$) 
as one of its solutions, it cannot however, be used to characterize the 
effect of scalar field models, in general, or modified gravity models on the 
observable Universe (\cite{review}). In this Section, we wish to determine the 
effectiveness of the topology data in constraining a second dark energy parameter. 
A popular two-parameter ansatz for $w(z)$ was introduced in (\cite{Chevallier:2000qy,linder-2003-90}) with the following form:
\be
w(z) = w_0 +w_a(1-a). 
\ee where $a = 1/(1+z)$.
This particular function is a favourite in the cosmology community because it 
solves the divergence problem at high redshift accompanying other parameterizations, 
but at the cost of a more rigid assumption of the behavior \emph{a priori} (\cite{Riess:2004nr}). 
It is also limited in how well it can cope with a rapidly evolving equation 
of state (\cite{Liddle:2006kn}). We simulate a set of genus values for a dark energy 
model with $w_0=-1$ and $w_a=0.5$ in three different 
binning schemes and add Gaussian noise.  Table \ref{tab_zbins_results} summarizes the results of the likelihood analysis 
when the simulated topology data in conjunction with the Planck priors is used to constrain 
$\bar{p}=\bra{\Omega_m, H_0, w_0, w_a}$. The constraints on $w_0$ weaken as expected to 
roughly $10\%$. We find that the topology data is relatively uninformative with regards to a 
second dark energy parameter, regardless of the number of redshift bins used.   
In fact, given that data is too noisy to detect any variation in $w$ with redshift, 
partitioning the data into redshift bins serves only to weaken the constraints on the average $w$.   
Figure \ref{figpriors_planck_priors_w0_wa} shows the $68\%$ and $95\%$ confidence regions in $\bra{w_0, w_a}$ 
when the data is binned into three redshift shells of equal volume and serves to illustrate the increase in the confidence region
accompanying the increased freedom in the dark energy model. Table \ref{tab_zbins_results} also indicates
that the use of a single redshift bin is prefered overall. 

\begin{table}
\begin{tabular}{|ccc|}
	\hline
Number of $z$ bins & $w_0$ & $w_a$ \\
\hline
Constant EOS: &$w_0=-0.9$  & \\
\\
1 & $-0.92^{+0.05}_{-0.05}$  & --  \\
2  & $-0.92^{+0.05}_{-0.05}$&  --\\
3  & $-0.92^{+0.04}_{-0.05}$& -- \\
Dynamical: & $w_0=-1$& $w_a=0.5$\\
\\
1 & $-1.04^{+0.1}_{-0.1}$ & $0.45^{+0.3}_{-0.5}$   \\
2 & $-0.93^{+0.1}_{-0.1}$ & $0.23^{+0.5}_{-0.6}$  \\
3  & $-0.98^{+0.1}_{-0.2}$ & $0.41^{+0.7}_{-0.6}$ \\ 
\hline
\end{tabular}
\label{tab_zbins_results}
\caption{Table summarizing the constraints on the dark energy EOS 
parameters for different binning schemes using the topology data and the 
Planck data. We assume the EOS to be constant from the last redshift bin 
at $z=0.6$ to the last scattering surface at $z=1089$. }
\end{table}

\section{Discussion}
\label{conclusion}
The question of the nature of dark energy is one of profound importance in cosmology today.   
The currently favoured model proposes that it is energy density associated with the vacuum, with a 
constant EOS of $w=-1$, identified as the mathematical equivalent of cosmological constant $\Lambda$ 
in the Einstein field equation.  
In the context of our current theories of structure formation, $\Lambda$ appears to be most successful 
in reproducing a wide range of present-day observations. 
However, it faces serious theoretical opposition \emph{and} several recent works 
have highlighted the biases towards $w=-1$ that could potentially arise from the 
inclusion of \emph{a prior} information when fitting the current data (\cite{bassett-2004-617, Linder:2004rm}). 
Future experiments offer the exciting prospect of determining whether the cosmological constant is 
indeed the correct model.  BOSS is forecasted to place constraints on $w$ at the
level of a few percent using clustering of LRGs on the scale of $100$ Mpc.  \cite{PK} proposed a 
way to use the nature of LRG clustering on a wider range of scales as measured by the 3D 
topology to probe the expansion rate.   Because genus statistics relate to the overall 
shape of the power spectrum as opposed to a single scale, they are robust against fortuitous noise measurements. Furthermore,
the toplogy measurements offer an alternative use of the data from redshift surveys and may be used as a cross check of the conclusions drawn 
from other techniques, such as the BAO method which uses the oscillations in the matter power spectrum to extract information. 

In this paper, we have evaluated the constraints on $w(z)$ that are achievable from the BOSS survey using the method presented in (\cite{PK}). 
We found that the BOSS survey using the topology method alone is capable of placing constraints of $5\%$ on a constant equation of state. 
Allowing for the possibility of time variation in $w(z)$ degrades the constraints on $w_0$ to $12\%$, while providing
weak evidence for a second dark energy parameter. Although the topology measurement of BOSS may not be capable of 
testing for dynamical behaviour in $w(z)$, it has been shown to provide a robust measurement of the average 
equation of state, which may be sufficient to rule out non-$\Lambda$ models. 

\section{Acknowledgements}
The authors would like to Changbom Park for suggesting this collaboration and 
Young-Rae Kim for providing the theoretical curves in Fig. \ref{g} based on the work in \cite{PK}.  
Thanks are due to Juhan Kim for providing the initial
density field of the Horizon Run simulation.  We also acknowledge the
Korean Institute for Advanced Study for providing computing resources
(KIAS linux cluster system, QUEST) for this work. 
CZ acknowledges support from the NRF (South Africa) and the PIRE grant (NSF).

\end{document}